\documentclass{ws-ijmpd}
\usepackage[super]{cite}
\usepackage{graphicx}
\begin{document}

\markboth{H. S. VIEIRA}
{Resonant frequencies of the hydrodynamic vortex}

\catchline{}{}{}{}{}

\title{Resonant frequencies of the hydrodynamic vortex}

\author{\footnotesize H. S. VIEIRA}
\address{Departamento de F\'{i}sica, Universidade Federal da Para\'{i}ba, Caixa Postal 5008, CEP 58051-970, Jo\~{a}o Pessoa, Para\'{i}ba, Brazil\\
Centro de Ci\^{e}ncias, Tecnologia e Sa\'{u}de, Universidade Estadual da Para\'{i}ba, CEP 58233-000, Araruna, Para\'{i}ba, Brazil\\
horacio.santana.vieira@hotmail.com}

\maketitle

\begin{history}
\received{Day Month Year}
\revised{Day Month Year}
\end{history}

\begin{abstract}
We study the sound perturbation of the hydrodynamic vortex geometry and present an exact expression for the resonant frequencies (quasispectrum) of this geometry. Exact solution for the radial part of the covariant Klein-Gordon equation in this spacetime is obtained, and is given in terms of the double confluent Heun functions. We found that the resonant frequencies are complex number.

\keywords{Vortex fluid flow; massless Klein-Gordon equation; double confluent Heun function; boundary value problem}
\end{abstract}

\ccode{PACS Nos.: 02.30.Gp, 04.20.Jb, 04.30.Nk, 04.70.-s, 43.20.+g, 47.35.Rs}
%
%
\section{Introduction}
The general theory of relativity has many sucessful predictions, in particular the existence of the most mysterious structures in the universe: black holes. In this sense the theoretical area of black hole physics became an attractive and productive line of research in the beginning of the last century. The knowledge of the behavior of different fields which interact with the gravitational field of black holes, certainly, will give us some relevant informations about the physics of these objects. In particular, the scalar field constitutes one of these fields which may give us important informations about the physics of a black hole \cite{ClassQuantumGrav.22.3833,PhysRevA.73.033604}.

With the aim of constructing a theory combining quantum mechanics and general relativity, the generalization of quantum field theory to curved spacetimes as well as their consequences have been discussed in the literature \cite{JMathPhys.5.848,Fulling:1989,Birrell:1994,Wald:1994,arXiv:gr-qc/0308048,ClassQuantumGrav.31.045003,AnnPhys.350.14,EurophysLett.109.60006,AnnPhys.362.576}. In fact, the experimental tests are difficult to make at the level of terrestrial laboratories. This opens the interest in analogue models that mimic the properties of astrophysics objects. In this sense, the field of analogue gravity allows in principle that the most processes of quantum field theory in curved spacetime can be studied in a laboratory. In particular, the use of supersonic acoustics flows as an analogy to gravitating systems has received a growing attention \cite{PhysRevLett.46.1351,Cardoso:2013}.

Solutions to the massless Klein-Gordon equation in the spacetime of both three-dimensional rotating and four-dimensional (4D) canonical acoustic black holes were obtained by Vieira and Bezerra \cite{GenRelativGravit.48.88} and are valid for the whole spacetime. In this context, the general solutions are given in terms of the confluent Heun functions and the appropriate forms of the equations suitable to study the Hawking radiation were presented and their solutions given.

In this present paper, we obtain the exact solution of the covariant Klein-Gordon equation for a massless scalar field in the hydrodynamic vortex spacetime, valid in the whole space. Furthermore, this solution is valid for all frequencies of the particles, that is, $\omega > 0$. It is given in terms of solutions of the Heun equations \cite{Ronveaux:1995,Slavyanov:2000}.

A property of interest that plays an important role in black hole physics is the Quasinormal Modes (see Beyer \cite{CommunMathPhys.204.397} and references therein). Many authors have considered the quasispectrum of massive fields in several black hole spacetimes. The terminology quasinormal modes have been used by Simone and Will \cite{ClassQuantumGrav.9.963} and Fiziev \cite{ClassQuantumGrav.23.2447}. They determine the late-time evolution of fields in the exterior of the black hole. Indeed, this is a boundary value problem for linear ordinary second-order differential equations with singularities at the endpoints of the interval of consideration \cite{JPhysAMathGen.31.4249}. Quasispectrum is not to be confused with quasinormal modes or bound states \cite{AnnPhys.373.28}. In the last few years, great attention has been given to the QNMs because it is believed that these models shed light on the solution or understanding the fundamental problems in loop quantum gravity \cite{PhysRevD.82.084037}. In this work, we show that the hydrodynamic vortex, which is an effective spacetime without an event horizon, but with ergoregion, is unstable under linearized perturbations.

The organization of the paper is as follows: in Sec. 2, we introduce the metric that corresponds to the hydrodynamic vortex, in Sec. 3, we write down the covariant Klein-Gordon equation for a massless scalar field in the background under consideration, in Sec. 4 we determine the resonant frequencies, and the large damping limits. Finally, we conclude in Sec. 5.
%
%
\section{Hydrodynamic vortex background}
The acoustic metric appropriate to a draining bathtub model, the so-called rotating acoustic black hole, which is the analog black hole metric (2+1)-dimensional with Lorentzian signature, is given by \cite{ClassQuantumGrav.15.1767}
\begin{equation}
ds^{2}=-c^{2}\ dt^{2}+\left(dr-\frac{A}{r}dt\right)^{2}+\left(r\ d\phi-\frac{B}{r}dt\right)^{2}\ ,
\label{eq:metrica_draining_bathtub}
\end{equation}
where $c$ is the speed of sound and is constant throughout the fluid flow. When we restricted to $A=0$, that is, no radial flow and generalised to an anisotropic speed of sound, the metric for a model that describes an acoustic geometry surrounding physical vortices takes the form \cite{LivingRevRelativity.8.12}
\begin{equation}
ds^{2}=-c^{2}\left(1-\frac{B^{2}}{c^{2}r^{2}}\right)dt^{2}+dr^{2}+r^{2}\ d\phi^{2}-2B\ dt\ d\phi\ ,
\label{eq:metrica_hydrodynamic_vortex}
\end{equation}
being $B$ a constant related to the circulation of the fluid.

We can write the metric tensor of the hydrodynamic vortex spacetime as
\begin{eqnarray}
(g_{\sigma\tau})=\left(
\begin{array}{ccccc}
	-c^{2}+\frac{B^{2}}{r^{2}} & & 0 & & -B\\
	 & & & & \\
	0 & & 1 & & 0\\
	 & & & & \\
	-B & & 0 & & r^{2}
\end{array}
\right)\ ,
\label{eq:metrica_matriz_hydrodynamic_vortex}
\end{eqnarray}
from which we obtain
\begin{equation}
g \equiv \det(g_{\sigma\tau})=-c^{2}r^{2}\ .
\label{eq:g_metrica_hydrodynamic_vortex}
\end{equation}
Thus, the contravariant components of $g_{\sigma\tau}$ are given by
\begin{eqnarray}
(g^{\sigma\tau})=\left(
\begin{array}{ccccc}
	-\frac{1}{c^{2}} & & 0 & & -\frac{B}{c^{2}r^{2}}\\
	 & & & & \\
	0 & & 1 & & 0\\
	 & & & & \\
	-\frac{B}{c^{2}r^{2}} & & 0 & & -\frac{B^{2}}{c^{2}r^{4}}+\frac{1}{r^{2}}
\end{array}
\right)\ .
\label{eq:var_contra_metrica_hydrodynamic_vortex}
\end{eqnarray}
As the Kerr black hole in general relativity, when the $g_{00}$ vanishes forms the radius of the ergosphere, $r_{e}$, given by
\begin{equation}
r_{e}=\frac{|B|}{c}\ ,
\label{eq:ergosphere}
\end{equation}
which coincides with the circle at which the (absolute value of the) background flow velocity equals the speed of sound $c$. Note that this spacetime does not have an event horizon.
%
%
\section{Exact solutions of the covariant Klein-Gordon equation}
For a barotropic and inviscid fluid with irrotational flow (though possibly time dependent), the equation of motion for the velocity potential describing a sound wave is identical to that for a minimally coupled massless scalar field, that is, to covariant Klein-Gordon equation, which has the form
\begin{equation}
\left[\frac{1}{\sqrt{-g}}\partial_{\rho}\left(g^{\rho\sigma}\sqrt{-g}\partial_{\sigma}\right)\right]\Psi=0\ .
\label{eq:Klein-Gordon_cova_hydrodynamic_vortex}
\end{equation}

Then, the covariant Klein-Gordon equation in the spacetime of a hydrodynamic vortex given by Eq.~(\ref{eq:metrica_hydrodynamic_vortex}) can be written as
\begin{equation}
\left[-\frac{r}{c}\frac{\partial^{2}}{\partial t^{2}}+\frac{\partial}{\partial r}\left(cr\frac{\partial}{\partial r}\right)+\left(\frac{c}{r}-\frac{B^{2}}{cr^{3}}\right)\frac{\partial^{2}}{\partial\phi^{2}}-\frac{2B}{cr}\frac{\partial^{2}}{\partial t\ \partial\phi}\right]\Psi(\mathbf{r},t)=0\ .
\label{eq:mov_hydrodynamic_vortex}
\end{equation}
The spacetime under consideration is time independent, so the time dependence that solves Eq.~(\ref{eq:mov_hydrodynamic_vortex}) may be separated as $\mbox{e}^{-i \omega t}$, where $\omega$ is the energy of the particles in the units chosen and we assume that $\omega > 0$. Moreover, rotational invariance with respect to $\phi$ permits us use the cylindrical symmetry of the effective background metric to write the solution in $\phi$ as $\mbox{e}^{im\phi}$, where $m$ is an integer. Therefore, $\Psi(\mathbf{r},t)$ can be written as
\begin{equation}
\Psi(\mathbf{r},t)=\sum_{m=-\infty}^{+\infty} R_{m\omega}(r)\mbox{e}^{im\phi}\mbox{e}^{-i \omega t}\ .
\label{eq:separacao_variaveis_hydrodynamic_vortex}
\end{equation}
Substituting Eq.~(\ref{eq:separacao_variaveis_hydrodynamic_vortex}) into Eq.~(\ref{eq:mov_hydrodynamic_vortex}), we find that
\begin{equation}
\left[\frac{d}{dr}\left(cr\frac{d}{dr}\right)+\left(\frac{\omega^{2}r}{c}-\frac{cm^{2}}{r}+\frac{B^{2}m^{2}}{cr^{3}}-\frac{2Bm\omega}{cr}\right)\right]R_{m\omega}(r)=0\ .
\label{eq:mov_radial_1_hydrodynamic_vortex}
\end{equation}

In what follows, we will obtain the exact and general solution for the radial equation given by Eq.~(\ref{eq:mov_radial_1_hydrodynamic_vortex}).
%
%
\subsection{Radial equation}
We can write down Eq.~(\ref{eq:mov_radial_1_hydrodynamic_vortex}) as
\begin{equation}
\frac{d^{2}R_{m\omega}}{dr^{2}}+\frac{1}{r}\frac{dR_{m\omega}}{dr}+\frac{1}{r}\left[\frac{\omega ^2}{c^2}r-\left(\frac{2 B m \omega }{c^2}+m^2\right)\frac{1}{r}+\frac{B^2 m^2}{c^2}\frac{1}{r^{3}}\right]R_{m\omega}=0\ .
\label{eq:mov_radial_2_hydrodynamic_vortex}
\end{equation}
This equation has singularities at $x=0$ and $x=\infty$. The transformation of Eq.~(\ref{eq:mov_radial_2_hydrodynamic_vortex}) to a Heun-type equation is achieved by setting
\begin{equation}
z=\frac{r^{2}+1}{r^{2}-1}\ .
\label{eq:z_hydrodynamic_vortex}
\end{equation}
Thus, we can written Eq.~(\ref{eq:mov_radial_2_hydrodynamic_vortex}) as
\begin{eqnarray}
&& \frac{d^{2}R_{m\omega}}{dz^{2}}+\left[\frac{2 z^5-4 z^3+2 z}{\left(z+1\right)^{3}\left(z-1\right)^{3}}\right]\frac{dR_{m\omega}}{dz}\nonumber\\
&& + \left[\frac{\left(A_{1}+A_{2}+A_{3}\right)z^{2}+2\left(A_{1}-A_{3}\right)z+\left(A_{1}-A_{2}+A_{3}\right)}{\left(z+1\right)^{3}\left(z-1\right)^{3}}\right]R_{m\omega}=0\ ,\nonumber\\
\label{eq:mov_radial_z_hydrodynamic_vortex_3}
\end{eqnarray}
where the coefficients $A_{1}$, $A_{2}$, and $A_{3}$ are given by:
\begin{equation}
A_{1}=\frac{\omega ^2}{c^2}\ ;
\label{eq:A1_hydrodynamic_vortex_3}
\end{equation}
\begin{equation}
A_{2}=-\left(\frac{2 B m \omega }{c^2}+m^2\right)\ ;
\label{eq:A2_hydrodynamic_vortex_3}
\end{equation}
\begin{equation}
A_{3}=\frac{B^2 m^2}{c^2}\ .
\label{eq:A3_hydrodynamic_vortex_3}
\end{equation}
Equation (\ref{eq:mov_radial_z_hydrodynamic_vortex_3}) is a special case of the double confluent Heun equation \cite{Annales.92.53} given by
\begin{eqnarray}
&& \frac{d^{2}U}{dz^{2}}+\left[\frac{2 z^5-\alpha  z^4-4 z^3+2 z+\alpha}{(z+1)^3(z-1)^3 }\right]\frac{dU}{dz}\nonumber\\
&& + \left[\frac{\beta  z^2- (-2 \alpha -\gamma )z+\delta}{(z+1)^3(z-1)^3 }\right]U=0\ ,
\label{eq:Heun_duplamente_confluente_forma_canonica}
\end{eqnarray}
where $U(z)=\mbox{HeunD}(\alpha,\beta,\gamma,\delta;z)$ are the double confluent Heun functions, with the parameters $\alpha$, $\beta$, $\gamma$, and $\delta$ according to the standard package of \textbf{Maple}\texttrademark \textbf{17}.

Thus, the general solution of the radial part of the Klein-Gordon equation for a massless scalar particle, the phonons, in the spacetime of a hydrodynamic vortex, given by Eq.~(\ref{eq:mov_radial_z_hydrodynamic_vortex_3}), over the entire range $0 \leq z < \infty$, can be written as
\begin{equation}
R_{m\omega}(z)=N_{m\omega}\ \mbox{HeunD}(\alpha,\beta,\gamma,\delta;z)\ ,
\label{eq:solucao_geral_radial_hydrodynamic_vortex}
\end{equation}
where $N_{m\omega}$ is a normalization constant to be determined, and the parameters $\alpha$, $\beta$, $\gamma$, and $\delta$ are now given by:
\begin{equation*}
\alpha=0\ ;
\label{eq:alpha_mov_radial_1_hydrodynamic_vortex_heun}
\end{equation*}
\begin{equation*}
\beta=A_{1}+A_{2}+A_{3}=\frac{B^2 m^2-2 B m \omega -c^2 m^2+\omega ^2}{c^2}\ ;
\label{eq:beta_mov_radial_1_hydrodynamic_vortex_heun}
\end{equation*}
\begin{equation*}
\gamma=2\left(A_{1}-A_{3}\right)=\frac{2 \omega ^2-2 B^2 m^2}{c^2}\ ;
\label{eq:gamma_mov_radial_1_hydrodynamic_vortex_heun}
\end{equation*}
\begin{equation*}
\delta=A_{1}-A_{2}+A_{3}=\frac{B^2 m^2+2 B m \omega +c^2 m^2+\omega ^2}{c^2}\ .
\label{eq:delta_mov_radial_1_hydrodynamic_vortex_heun}
\end{equation*}

If we consider a standard power series expansion around the origin (a regular point) of the double confluent Heun function, and taking into account the presence of two irregular singularities located at $-1$ and $1$, with radius of convergence of this series being $|z| < 1$, we can write
\begin{equation*}
\mbox{HeunD}(\alpha,\beta,\gamma,\delta;z)=1+\frac{1}{2}\delta\ z^{2}+\left(\frac{1}{6}\delta\alpha+\frac{1}{3}\alpha+\frac{1}{6}\gamma\right)z^{3}+\ldots\ .
\label{eq:serie_HeunD_todo_z}
\end{equation*}
%
%
\section{Resonant frequencies}
In order to compute the resonant frequencies, we need to impose boundary conditions on the solutions at the asymptotic region (infinity), which in this case, requires the necessary condition for a polynomial form of $R_{m\omega}(z)$ (for a review, see \cite{ModPhysLettA.21.1601} and references therein). The solution should be finite on the horizon and well behaved far from the black hole \cite{AnnPhys.373.28}.

The asymptotic and numeric study of the boundary problems related to the double confluent case of Heun's differential equation was carried out by Lay et al. \cite{JPhysAMathGen.31.8521}. However, to do this we will follow the method developed by Gurappa and Panigrahi \cite{JPhysAMathGen.37.L605}, because it is the most suitable for the double confluent Heun equation in the canonical form given by Eq.~(\ref{eq:Heun_duplamente_confluente_forma_canonica}) \cite{RomJournPhys.59.224}. Therefore, in general, for a single-variable differential equation which can be written as
\begin{equation}
[F(D)+P(z,d/dz)]R_{m\omega}(z)=0\ ,
\label{eq:form_generally}
\end{equation}
where
\begin{equation}
D \equiv z\frac{d}{dz}\ ,
\label{eq:D}
\end{equation}
and
\begin{equation}
F(D)=\sum_{n}a_{n}D^{n}
\label{eq:F(D)}
\end{equation}
is a diagonal operator in the space of monomials, with $a_{n}$ being the coefficients of expansion, and $P(z,d/dz)$ is an arbitrary polynomial function of $z$ and $d/dz$ (except $D$), one has to impose the condition
\begin{equation}
F(D)z^{n}=0\ .
\label{eq:condition}
\end{equation}

\begin{figure}[htbp]
	\centering
		\includegraphics[scale=0.60]{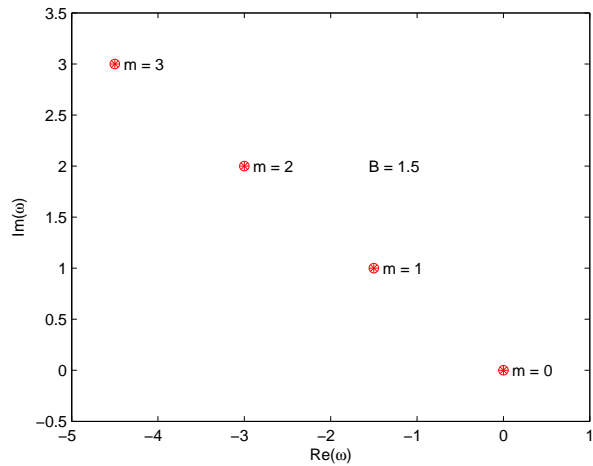}
	\caption{Comparison of the characteristic complex resonant frequencies of the fundamental mode ($n=0$). We focus on azimuthal numbers $m=0,1,2,3$ and circulation $B=1.5$. The unit $c\equiv 1$ was chosen.}
	\label{fig:n=0_hyd_vor}
\end{figure}

Rewriting Eq.~(\ref{eq:mov_radial_z_hydrodynamic_vortex_3}) after multiplying it by $[(z+1)(z-1)]^{3}$, we get
\begin{eqnarray}
&& \biggl[3D^{2}-D+(A_{1}-A_{2}+A_{3})+(A_{1}+A_{2}+A_{3})z^{2}+2(A_{1}-A_{3})z\nonumber\\
&& +(-1-3z^{4}+z^{6})\frac{d^{2}}{dz^{2}}+(-4z^{3}+2z^{5})\frac{d}{dz}\biggr]R_{m\omega}(z)=0\ .
\label{eq:mov_radial_z_hydrodynamic_vortex_4}
\end{eqnarray}
In the above equation, it turns out that the diagonal operator, $F(D)$, and the polynomial function, $P(z,d/dz)$, are given by
\begin{equation}
F(D)=3D^{2}-D+\left(A_{1}-A_{2}+A_{3}\right)\ ,
\label{eq:essential_operator}
\end{equation}
\begin{eqnarray}
P(z,d/dz) & = & (A_{1}+A_{2}+A_{3})z^{2}+2(A_{1}-A_{3})z\nonumber\\
& + & (-1-3z^{4}+z^{6})\frac{d^{2}}{dz^{2}}+(-4z^{3}+2z^{5})\frac{d}{dz}\ .
\label{eq:P}
\end{eqnarray}
Now, the condition (\ref{eq:condition}) lead directly to an exact determination of the resonant frequencies as follows:
\begin{equation}
\omega_{n}(m,B)=-Bm \pm ic\sqrt{m^{2}+3n^{2}-n}\ ,
\label{eq:energy_levels}
\end{equation}
where $n=0,1,2,\ldots$. This is a nontrivial quantization law, because it has the properties being a complex number and independent of the radius of the ergosphere of the hydrodynamic vortex (see Fig.~\ref{fig:n=0_hyd_vor} and Table~\ref{tab:QNMs_frequencies}).

It is worth calling attention to the fact that, from Eq.(\ref{eq:energy_levels}), the hydrodynamic vortex has the following symmetries
\begin{equation}
\omega_{n}(m,B)=-\omega_{n}^{*}(-m,B)=-\omega_{n}^{*}(m,-B)=\omega_{n}(-m,-B)\ ,
\label{eq:energy_levels_symmetries}
\end{equation}
where $``^{*}"$ denotes complex conjugation.

Following the standard convention to order the resonant frequencies frequencies by their imaginary part \cite{ClassQuantumGrav.26.163001}, we have that the largest imaginary component $\mbox{Im}(\omega_{n})$ corresponds to the fundamental mode. Then, if $\mbox{Im}(\omega_{n}) > 0$ the mode is unstable, so that the fundamental mode corresponds to the smallest instability time scale. On the other hand, if $\mbox{Im}(\omega_{n}) < 0$ the mode is stable, and the fundamental mode corresponds to the longest-lived mode.

\begin{table}[htbp]
\tbl{Resonant frequencies of the modes $n=0,1,2$ for $m=0,1,2,3$ and circulation $B=0.5$. The unit $c \equiv 1$ was chosen.}
		{\begin{tabular}{ccccccc}\hline
			& \multicolumn{2}{c}{$n=0$} & \multicolumn{2}{c}{$n=1$} & \multicolumn{2}{c}{$n=2$} \\
			$m$ & $\mbox{Re}(\omega)$ & $\mbox{Im}(\omega)$ & $\mbox{Re}(\omega)$ & $\mbox{Im}(\omega)$ & $\mbox{Re}(\omega)$ & $\mbox{Im}(\omega)$ \\\hline
			0 & \ 0.0 & 0.0 & \ 0.0 & 1.4 & \ 0.0 & 3.2 \\
			1 & -0.5 & 1.0 & -0.5 & 1.7 & -0.5 & 3.3 \\
			2 & -1.0 & 2.0 & -1.0 & 2.4 & -1.0 & 3.7 \\
			3 & -1.5 & 3.0 & -1.5 & 3.3 & -1.5 & 4.4 \\\hline
		\end{tabular}
	\label{tab:QNMs_frequencies}}
\end{table}

In the $n \rightarrow \infty$ limit, that is, the infinite limit damping, we have
\begin{equation}
\omega_{\infty}=-Bm \pm i\left(n-\frac{1}{6}\right)\sqrt{3c^{2}}\ .
\label{eq:energy_levels_infinity}
\end{equation}
Therefore, those results show the instabilities of the hydrodynamic vortex under sonic perturbations both in the fundamental mode and in the limit that the background fluid smoothly exceeding the velocity of sound.
%
%
\section{Conclusions}
In this work we have presented analytic solution of the radial part of the covariant Klein-Gordon equation for phonons (acoustic massless scalar particle) in the hydrodynamic vortex spacetime.

This general solution is analytic for all spacetime, which means, in the region between the ergosphere and infinity. The radial solution is given in terms of the double confluent Heun functions, and is valid over the range $0 \leq z < \infty$.

We have computed the exact values of the resonant frequencies of the hydrodynamic vortex, which are complex number. The resonant frequencies formula show that the instabilities of the hydrodynamic vortex under sonic perturbations both in the fundamental mode and in the limit that the background fluid smoothly exceeding the velocity of sound.

We also want to note that the resonant frequencies do not depend on the radius of the ergosphere, but depend on the angular momentum of the perturbation.
%
%
\section*{Acknowledgments}
The author would like to thank Prof. V. B. Bezerra for the fruitful discussions and to thank also Conselho Nacional de Desenvolvimento Cient\'{i}fico e Tecnol\'{o}gico (CNPq) for financial support. H. S. V. is funded through the research Project No. 140612/2014-9.
%
%

%
%

\begin{thebibliography}{0}
\bibitem{ClassQuantumGrav.22.3833} T. R. Slatyer and C. M. Savage, \textit{Class. Quantum Grav.} \textbf{22}, 3833 (2005).
\bibitem{PhysRevA.73.033604} F. Federici, C. Cherubini, S. Succi and M. P. Tosi, \textit{Phys. Rev. A} \textbf{73}, 033604 (2006).
%
\bibitem{JMathPhys.5.848} R. Haag and D. Kastler, \textit{J. Math. Phys.} \textbf{5}, 848 (1964).
\bibitem{Fulling:1989} S. A. Fulling, \textit{Aspects of quantum field theory in curved space-time}, (Cambridge University Press, New York, 1989).
\bibitem{Birrell:1994} N. D. Birrell and P. C. W. Davis, \textit{Quantum fields in curved space}, (Cambridge University Press, New York, 1994).
\bibitem{Wald:1994} R. M. Wald, \textit{Quantum field theory in curved spacetime and black hole thermodynamics}, (Chicago Lectures in Physics, Chicago, 1994).
\bibitem{arXiv:gr-qc/0308048} T. Jacobson, \textit{Introduction to quantum fields in curved spacetime and the Hawking effect}, \textit{arXiv:gr-qc/0308048} \textbf{[gr-qc]} (2003).
\bibitem{ClassQuantumGrav.31.045003} V. B. Bezerra, H. S. Vieira and A. A. Costa, \textit{Class. Quantum Grav.} \textbf{31}, 045003 (2014).
\bibitem{AnnPhys.350.14} H. S. Vieira, V. B. Bezerra and C. R. Muniz, \textit{Ann. Phys. (NY)} \textbf{350}, 14 (2014).
\bibitem{EurophysLett.109.60006} H. S. Vieira, V. B. Bezerra and A. A. Costa, \textit{Europhys. Lett.} \textbf{109}, 60006 (2015).
\bibitem{AnnPhys.362.576} H. S. Vieira, V. B. Bezerra and G. V. Silva, \textit{Ann. Phys. (NY)} \textbf{362}, 576 (2015).
%
\bibitem{PhysRevLett.46.1351} W. G. Unruh, \textit{Phys. Rev. Lett.} \textbf{46}, 1351 (1981).
\bibitem{Cardoso:2013} V. M. S. Cardoso, L. C. B. Crispino, S. Liberati, E. S. Oliveira and M. Visser, \textit{Analogue spacetimes: The first thirty years}, (Editora Livraria da F\'{i}sica, S\~{a}o Paulo, 2013).
%
\bibitem{GenRelativGravit.48.88} H. S. Vieira and V. B. Bezerra, \textit{Gen. Relativ. Gravit.} \textbf{48}, 88 (2016).
%
\bibitem{Ronveaux:1995} A. Ronveaux, \textit{Heun's differential equations}, (Oxford University Press, New York, 1995).
\bibitem{Slavyanov:2000} S. Y. Slavyanov and W. Lay, \textit{Special functions}, (Oxford University Press, New York, 2000).
%
\bibitem{CommunMathPhys.204.397} H. R. Beyer, \textit{Commun. Math. Phys.} \textbf{204}, 397 (1999).
%
\bibitem{ClassQuantumGrav.9.963} L. E. Simone and C. M. Will, \textit{Class. Quantum Grav.} \textbf{9}, 963 (1992)
\bibitem{ClassQuantumGrav.23.2447} P. P. Fiziev, \textit{Class. Quantum Grav.} \textbf{23}, 2447 (2006)
%
\bibitem{JPhysAMathGen.31.4249} W. Lay and S. Y. Slavyanov, \textit{J. Phys. A: Math. Gen.} \textbf{31}, 4249 (1998).
%
\bibitem{AnnPhys.373.28} H. S. Vieira and V. B. Bezerra, \textit{Ann. Phys. (NY)} \textbf{373}, 28 (2016).
%
\bibitem{PhysRevD.82.084037} S. R. Dolan, L. A. Oliveira and L. C. B. Crispino, \textit{Phys. Rev. D} \textbf{82}, 084037 (2010).
%
\bibitem{ClassQuantumGrav.15.1767} M. Visser, \textit{Class. Quantum Grav.} \textbf{15}, 1767 (1998).
%
\bibitem{LivingRevRelativity.8.12} C. Barcel\'{o}, S. Liberati and M. Visser, \textit{Living Rev. Relativity} \textbf{8}, 12 (2005).
%
\bibitem{Annales.92.53} A. Decarreau, M. C. D. Lepage, P. Maroni, A. Robert and A. Ronveaux, \textit{Ann. Soc. Sci. Bruxelles} \textbf{92}, 53 (1978).
%
\bibitem{ModPhysLettA.21.1601} J. Saavedra, \textit{Mod. Phys. Lett. A} \textbf{21}, 1601 (2006).
%
\bibitem{JPhysAMathGen.31.8521} W. Lay, K. Bay and S. Y. Slavyanov, \textit{J. Phys. A: Math. Gen.} \textbf{31}, 8521 (1998).
%
\bibitem{JPhysAMathGen.37.L605} N. Gurappa and P. K. Panigrahi, \textit{J. Phys. A: Math. Gen.} \textbf{37}, L605 (2004).
%
\bibitem{RomJournPhys.59.224} M. A. Dariescu, D. Mihu and C. Dariescu, \textit{Rom. Journ. Phys.} \textbf{59}, 224 (2014).
%
\bibitem{ClassQuantumGrav.26.163001} E. Berti, V. Cardoso and A. O. Starinets, \textit{Class. Quantum Grav.} \textbf{26}, 163001 (2009).
\end{thebibliography}
\end{document}